\documentclass[twocolumn,fp]{jpsj3}
\usepackage{txfonts}
\usepackage{graphicx}
\usepackage{amssymb}
\usepackage{dcolumn}
\title{Ground state and finite temperature behavior 
of $\frac{1}{4}$-filled band zigzag ladders}

\author{R. Torsten Clay$^{1,2}$\thanks{E-mail address: r.t.clay@msstate.edu}, 
Jeong-Pil Song$^{1}$, Saurabh Dayal$^{1}$, and Sumit Mazumdar$^{3}$}
\inst{$^{1}$Department of Physics and Astronomy and HPC$^2$ Center for Computational 
Sciences, Mississippi State University, Mississippi State, MS, 39762 \\
$^{2}$Institute for Solid State Physics, The University of Tokyo,
Kashiwa 277-8581, Japan \\
$^{3}$Department of Physics, University of Arizona, Tucson, AZ, 85721}

\abst{
We consider the simplest example of lattice frustration in the
$\frac{1}{4}$-filled band, a one-dimensional chain with next-nearest
neighbor interactions. For this zigzag ladder with electron-electron
as well as electron-phonon interactions we present numerical results
for ground state as well as thermodynamic properties. In this system
the ground state bond distortion pattern is independent of
electron-electron interaction strength. The spin gap from the ground
state of the zigzag ladder increases with the degree of frustration.
Unlike in one-dimension, where the spin-gap and charge ordering
transitions can be distinct, we show that in the ladder they occur
simultaneously. We discuss spin gap and charge ordering transitions in
$\frac{1}{4}$-filled materials with one, two, or three dimensional
crystal structures. We show empirically that regardless of
dimensionality the occurrence of simultaneous or distinct charge and
magnetic transitions can be correlated with the ground state bond
distortion pattern.
}
\kword{ladder, spin gap, lattice frustration, 1/4 filling, organic conductors}

\begin{document}
\maketitle
\def\newblock{\hskip .11em plus .33em minus .07em}

\section{Introduction}
\label{intro}

The consequences of geometric lattice frustration on ground and
excited state behavior of spin systems have been widely studied over
the past two decades \cite{Diep,Lacroix}.  Even as many of the issues
are still being debated for the complex two- and three-dimensional (2D
and 3D) lattices, consensus on some of the simpler models have been
reached.  One such model frustrated system is the spin-$\frac{1}{2}$
zigzag ladder, with nonzero antiferromagnetic exchanges $J_1$ and
$J_2$ between nearest and next-nearest neighbor spins. The ground
state here is a valence bond solid (VBS) for $J_2$ = 0.5$J_1$
\cite{Majumdar69a,Majumdar69b}, and is dimerized for $J_2 \geq
0.2411J_1$ \cite{Okamoto92a,Chitra95a}, where the excitations are spin
solitons \cite{Shastry81a,Sorensen98a}.  Some theoretical results
exist also for $J_2 > J_1$ \cite{White96a,Itoi01a,Kumar10a}.

Spin Hamiltonians are restricted to a charge per site $\rho=1$
($\frac{1}{2}$-filled band) and are obtained in the limit of very
large onsite Hubbard repulsion $U$ between electrons or holes where
there are no charge degrees of freedom.  The literature on frustrated
non-$\frac{1}{2}$-filled bands, where the charge degrees of freedom
are non-vanishing even for $U \to \infty$, remains relatively
sparse. Much of the literature on non-$\frac{1}{2}$-filled band
frustrated systems is for $\rho=\frac{1}{2}$, or the
$\frac{1}{4}$-filled band, which is of interest both because under
certain conditions it can be described within an effective
$\frac{1}{2}$-filled band picture \cite{Kino95a,Kanoda06a,Powell11a},
and because there exist many $\frac{1}{4}$-filled band frustrated
materials, both organic and inorganic, that exhibit novel behavior
including charge, spin and orbital-ordering, and superconductivity
(see below). We have recently shown that there is a strong tendency to
form local spin-singlets in the $\frac{1}{4}$-filled band with strong
electron-electron interactions, in both one dimension (1D) and 2D, 
and especially in the presence of lattice
frustration, which enhances quantum effects \cite{Li10a,Dayal11a}.
The frustration-driven singlet formation does not occur if only the
frustrating Coulomb interactions are included \cite{Merino05a}; the
inclusion of the frustrating electron-hopping integral, with or
without the corresponding Coulomb interactions is essential. Singlet
formation in the $\frac{1}{4}$-filled band is accompanied by
charge-ordering (CO), leading to what we have termed as a {\it
  paired-electron crystal} (PEC), which consists of pairs of
singly-occupied sites separated by pairs of vacant sites. The PEC is
the $\frac{1}{4}$-filled band equivalent of the VBS. The difference
from the standard VBS lies in the possibility that the PEC gives way
to a {\it paired-electron liquid} under appropriate conditions, which
might then lead to superconductivity. Clearly it is desirable to
extend of our current work on the ground state of the PEC to excited
states.
\begin{figure}
\centerline{\resizebox{3.2in}{!}{\includegraphics{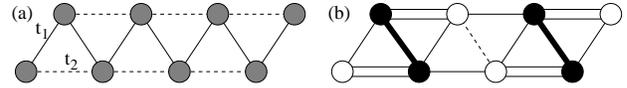}}}
\caption{Zigzag ladder lattice and ground state configurations for
  density $\rho=\frac{1}{2}$.  (a) The ground state for
  $t_2/t_1>1.707$ has uniform charges and bonds. (b) For
  $t_2/t_1<1.707$, the ground state is a PEC with coexisting charge
  and bond order. Filled (open) circles correspond to sites with large
  (small) charge density.  The heavy line indicates the strongest
  (spin-singlet) bond.  Double, single, and dotted lines indicate
  strong, intermediate, and weak bonds, respectively.}
\label{lattice}
\end{figure}

In the present paper, we discuss theoretical results for the PEC that
occurs in the simplest $\frac{1}{4}$-filled frustrated lattice with
interacting electrons. The lattice geometry here is the same as in the
spin zigzag ladder (see Fig.~\ref{lattice}(a)), with nearest and next
nearest neighbor electron hopping matrix elements $t_1$ and $t_2$,
respectively. In our previous work on the ground state of the
$\frac{1}{4}$-filled band zigzag ladder \cite{Clay05a}, we had shown
that in the presence of electron-lattice coupling (either nearest
neighbor along the zigzag direction, or second neighbor, or both), and for
$t_2/t_1$ less than a critical value $(t_2/t_1)_c$ (see below) there
occurs a coupled bond-charge density wave (BCDW), as shown in
Fig.~\ref{lattice}(b). Alternate rungs of the zigzag ladder are
occupied by singlet spin-coupled charge-rich sites, and charge-poor
sites occupy the other set of rungs. The CO pattern here characterizes
the BCDW as a PEC.  We discuss below the nature of the ground state in
this system for the complete parameter range $t_2/t_1 \leq
(t_2/t_1)_c$.  We also present numerical results for spin excitations
and thermodynamics for the $\frac{1}{4}$-filled band zigzag ladder
within the interacting electron Hamiltonian and discuss applications
of the model to real systems.

Although our actual work is limited to the zigzag ladder, our analysis
leads to an empirical understanding of a perplexing observation in the
$\frac{1}{4}$-filled band systems with interacting electrons in
general. In many such systems, irrespective of dimensionality, there
often are two transitions, (i) a high temperature metal-insulator (MI)
transition at T$_{\rm{MI}}$ that is accompanied by CO without any
perceptible effect on the magnetic behavior, and (ii) a second
insulator-insulator transition at a lower temperature T$_{\rm{SG}}$
where a spin gap (SG), seen in the magnetic susceptibility, occurs.
In a second class of systems with nearly identical chemical
constituents there occurs however a single transition with
T$_{\rm{MI}}$ = T$_{\rm{SG}}$. We point out a well-defined correlation
between this ``two-versus-one'' transition and the pattern of the bond
distortion in the spin-gapped phase.  Furthermore, many interacting
$\frac{1}{4}$-filled band materials are superconducting, often under
pressure.  There appears to exist a correlation also between
superconductivity and the bond distortion pattern at the lowest
temperatures.

The organization of the paper is as follows. In Section \ref{model} we
introduce our Hamiltonian for the $\frac{1}{4}$-filled band zigzag
ladder.  In Section \ref{summary} we present a brief summary of the
earlier results for the 1D limit of this model
\cite{Ung94a,Kuwabara03a,Clay03a,Clay07a,Yoshimi11a}, 
and for the ground state of the zigzag
ladder \cite{Clay05a}. Two different bond distortion patterns,
accompanied however with the same charge distortion pattern are
possible in the 1D limit. We point out the two bond distortion
patterns correspond to two different mappings of the
$\frac{1}{4}$-filled band to effective $\frac{1}{2}$-filled bands. In
the 1D limit both mappings are valid. In contrast, only one mapping is
applicable to the zigzag ladder, and there is a single unambiguous PEC
here.  The results of our numerical calculations for the zigzag ladder
are presented in Section \ref{results}, where we discuss both ground
state and temperature-dependent behavior. In Section \ref{discussion}
we discuss the implications of our results for real
$\frac{1}{4}$-filled materials. Appendix \ref{appendixa} contains details of the
numerical method, based on a matrix-product state (MPS)
representation, used for our large-lattice calculations.
Appendix \ref{appendixb} contains details of the
finite-size scaling of our ground-state calculations.

\section{Theoretical model and parameter space}
\label{model}
The Hamiltonian for the zigzag ladder is
\begin{eqnarray}
H&=&-t_1\sum_{i}(1+\alpha_1\Delta_{i,i+1})B_{i,i+1}
 +\frac{1}{2}K_1\sum_{i}\Delta^2_{i,i+1} \nonumber \\
&-& t_2\sum_{i}(1+\alpha_2\Delta_{i,i+2})B_{i,i+2} +\frac{1}{2}K_2\sum_i 
\Delta^2_{i,i+2} \nonumber \\
&+&  \sum_i(U n_{i\uparrow}n_{i\downarrow} + V_1 n_in_{i+1} +V_2  n_in_{i+2}).
\label{ham}
\end{eqnarray}
In Eq.~\ref{ham}
$B_{i,j}=\sum_\sigma(c^\dagger_{j,\sigma}c_{i,\sigma}+H.c.)$ is the
kinetic energy operator for the bond between sites $i$ and $j$, where
$c^\dagger_{i,\sigma}$ creates an electron of spin $\sigma$ on site
$i$. $n_{i\sigma}=c^\dagger_{i,\sigma}c_{i,\sigma}$ is the density
operator, and $n_i=n_{i\uparrow}+n_{i\downarrow}$.  $t_1$ and $t_2$
are hopping integrals along the zigzag rung direction and the rail
direction, respectively, as shown in Fig.~\ref{lattice}.  The lattice
may also be viewed as a 1D chain with nearest neighbor hopping $t_1$
and frustrating second-neighbor hopping $t_2$.  In our reference to
the 1D limit below, however, we will imply $t_2=0$ and $V_2=0$.  We
give energies in units of $t_1$ and fix $t_1$=1.  $\Delta_{i,j}$ is
the deformation of the bond between sites $i$ and $j$; $\alpha_1$ and
$\alpha_2$ are the inter-site electron-phonon (e-p) coupling constants
with corresponding spring constants $K_1$ and $K_2$, which for
simplicity we choose to be identical in the $t_1$ and $t_2$ directions
($\alpha_1=\alpha_2\equiv\alpha$ and $K_1=K_2\equiv K$).  For all
results below we fix $K_1=K_2=2$. We have omitted intra-site e-p
coupling terms in Eq.~\ref{ham}, as apart from changing the strength
of the charge order, they do not greatly change the thermodynamic
properties \cite{Clay07a}.  $U$ is the onsite Coulomb interaction and
$V_1$ and $V_2$ are the nearest-neighbor Coulomb interactions for
$t_1$ and $t_2$ bonds. We will consider only the case $V_1=V_2=V$. In
the 1D limit, the ground state is a Wigner crystal (WC) for
$V_1>V_1^c$, where $V_1^c=2|t_1|$ for $U \to \infty$ and is larger for
finite $U$.  Our discussions of the 1D limit are for $V_1<V_1^c$.  All
calculations use periodic boundary conditions.

\section{Summary of earlier results.}
\label{summary}
Given the complexity of our results, it is useful to have a brief
summary of our earlier results. While the numerical results for the 1D
limit and the ground state of the zigzag ladder have been presented
before, the mappings of the ground states to the effective
$\frac{1}{2}$-filled band model that we discuss below give new
insight.

\subsection{1D limit}
\label{1dsummary}

\begin{figure}
\centerline{\resizebox{3.0in}{!}{\includegraphics{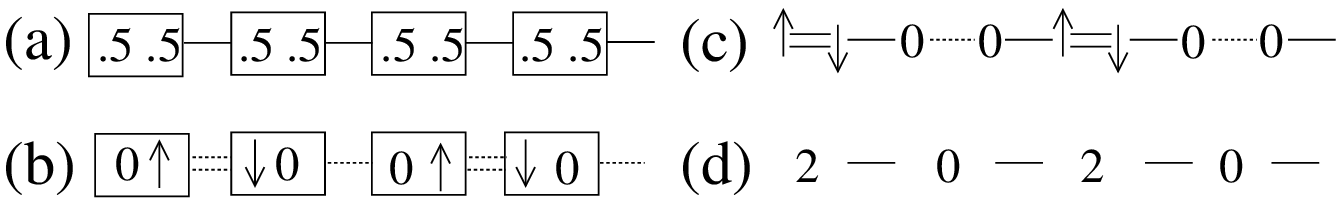}}}
\caption{
  (a) Bond-dimerized phase with uniform charge density. Boxes denote
  dimer units. (b) SP state evolving from (a), with bond pattern
  SWSW$^\prime$.  Strong `S' dimer bonds are represented by boxes
  enclosing two sites. Zeroes indicate sites with charge density
  0.5-$\delta$ while spins indicate sites with charge density
  0.5+$\delta$. Double dotted bonds are stronger than single-dotted
  bonds, but weaker than the dimer bonds.  (c) Bond pattern SMWM found
  in the case of relatively weaker e-e interactions.  Here double
  lines indicate the strongest `S' bonds, single lines medium strength
  `M' bonds, and dotted lines weak `W' bonds. (d) Effective
  $\frac{1}{2}$-filled band model for (c).}
\label{1d}
\end{figure}

For moderate to strong electron-electron (e-e) interactions but
$V_1<V_1^c$, the $\frac{1}{4}$-filled band is bond- or
charge-dimerized at T$<$T$_{\rm{MI}}$. In either case the ground state
enters a spin-Peierls (SP) phase \cite{Clay07a,Yoshimi11a} for T$<$T$_{\rm{SG}}$.
We show in Fig.~\ref{1d}(a) a schematic of the bond-dimerized phase,
with site charge densities 0.5, strong intradimer bonds and weak
interdimer bonds.  The dimer unit cells (the boxes in the figure)
containing a single electron can be thought of as effective sites,
which leads to the effective $\frac{1}{2}$-filled band description in
this case \cite{Powell11a}.  The description of the SP state is then
as shown in Fig.~\ref{1d}(b), with alternating {\it inter}-dimer
bonds. The overall pattern of bond strength here is
$\cdots$Strong-Weak-Strong-Weak$^\prime\cdots$ (SWSW$^\prime$)
\cite{Ung94a,Mazumdar00a,Clay03a}.  Unlike the true
$\frac{1}{2}$-filled band, however, the charge degrees of freedom
internal to the dimer unit cell are relevant in the
$\frac{1}{4}$-filled band, and the charge distribution is as shown in
the Fig.~\ref{1d}(b).  The coexisting charge order pattern is written
as $\cdots$1100$\cdots$, where `1' (`0') denotes site charge density
of 0.5+$\delta$ (0.5-$\delta$). The CO amplitude $\delta$ varies with
e-e and e-p interaction strengths. Nearest-neighbor spin-singlet
coupling between the electrons on charge-rich `1' sites that are
linked by the W$^\prime$ inter-dimer bond gives the spin gap of this
SP phase.  This coexisting broken symmetry state has been termed a
BCDW \cite{Ung94a,Mazumdar00a,Clay03a}, and is the simplest example of
the PEC found more generally in $\frac{1}{4}$-filled systems beyond
the 1D limit \cite{Li10a,Dayal11a}.  Finite temperature phase
transition to the SWSW$^\prime$ phase, starting from the uniform phase
and through the dimer phase of Fig.~\ref{1d}(a), has been demonstrated
within an extended Hubbard model that incorporated interchain
interactions at a mean-field level \cite{Otsuka08a}.

The pattern of bond order modulation in the 1D PEC is not unique but
depends on the strength of e-e interactions \cite{Ung94a}. The
bond-order pattern that dominates for relatively weaker e-e
interactions, or for relatively strong e-p interaction has the form
$\cdots$Strong-Medium-Weak-Medium$\cdots$ (SMWM) \cite{Ung94a}.
Importantly, in this bond pattern the coexisting charge modulation
pattern is again $\cdots$1100$\cdots$ but the spin singlets now
coincide with the strongest `S' {\it intra}-dimer bonds
(Fig.~\ref{1d}(c)).  As shown in Fig.~\ref{1d}(d), the corresponding
effective $\frac{1}{2}$-filled band is now different; pairs of sites
with large (small) charge densities are now mapped onto effective
sites with charge density 2 (0). The bonds between the effective sites
are now equivalent, and the effective state is now a
$\frac{1}{2}$-filled band {\it site}-diagonal CDW, which is known to
have a single transition where charge and magnetic gaps open
simultaneously.  Further, as the strongest bond coincides with the
location of the spin-singlet, this state is also expected to have a
larger SG and transition temperature than T$_{\rm{SG}}$ in the
SWSW$^\prime$ case. A simple physical picture for two-versus-one
transition is thus obtained from the 1D study: if the strongest bonds
are between a charge-rich and a charge-poor site, the spin-singlet
formed are inter-dimer and there will occur two transitions; if,
however, the strongest bond is a 1-1 bond, the spin singlet is located
on an intra-dimer bond and there is a single transition.

\subsection {Zigzag ladder}

The energy dispersion relation in the zigzag ladder in the
noninteracting limit is a sum of two terms,
\begin{equation}
E(k)=-2t_1\cos(q)-2t_2\cos(2q).
\label{bands}
\end{equation}
Here $q$ refers to a wavenumber along the $t_1$ direction, viewing the
ladder as a 1D chain with second-neighbor interactions. The topology
of the bandstructure changes at
$t_2/t_1=(t_2/t_1)_c=(2+\sqrt{2})/2=1.707\cdots$; for
$t_2/t_1<(t_2/t_1)_c$ the Fermi surface consists of two points at
$q=k_F=\frac{\pi}{4}$, while for $t_2/t_1>(t_2/t_1)_c$ there are four
such points. In the presence of e-p coupling the ground state is then
unstable\cite{Clay05a} to a Peierls distorted state for
$t_2/t_1<(t_2/t_1)_c$.  As in 1D, the ground state in the distorted
region again has $\cdots1100\cdots$ CO and a spin gap.

Importantly, for the $\rho=\frac{1}{2}$ zigzag ladder, there is no WC
state with a charge order {\it distinct} from the $\cdots1100\cdots$
charge order found in the PEC (for $V_1 \sim V_2$).  In the zigzag
ladder, the WC charge pattern $\cdots1010\cdots$ CO can be placed
along the two $t_2$ directions in two different ways, both of which
lead to the same PEC state shown in Fig.~\ref{lattice}(b).  Placing
the WC CO pattern $\cdots1010\cdots$ along the zigzag direction would
place all charge on a single $t_2$ chain and is stable only in the
limit $V_1\gg V_2$.  Thus only the PEC of Fig.~\ref{lattice}(b) is
stable for realistic $V_1 \sim V_2$. As we show below, this has
important consequences for both the number of transitions expected in
the zigzag ladder and the nature of the spinon excitations.

\section{Numerical Results for the zigzag ladder}
\label{results}

We use two different numerical methods to solve Eq.~\ref{ham}.  For
the ground state order parameters and the spin gaps in the interacting
case we use a new variational quantum Monte Carlo (QMC) method using a
MPS basis \cite{Sandvik07a}. For quasi-1D systems, this MPS-QMC method
provides accuracy similar to that of the Density Matrix
Renormalization Group (DMRG)\cite{White92a,White93a} method.  Similar
methods using MPS representations have predominantly been used to
study quantum spin systems. Our results here show however that they
may be successfully applied to electronic models as well.  Details on
the application of this method to Hubbard-type models are discussed in
Appendix \ref{appendixa}. For finite temperatures our calculations are
for zero e-p interactions ($\alpha_1=\alpha_2=0$).  This is because
information about the tendency to distortion exists in the
wavefunction even without inclusion of explicit e-p interactions
\cite{Hirsch84a}.  We use here the standard determinantal QMC method
\cite{Loh92a}. While the lowest temperatures achievable by this method
are limited by the Fermion sign problem, in the present system for
density $\rho=\frac{1}{2}$, inverse temperatures of $\beta \approx
10-16$ are reachable for parameters $U\approx 4$ and $V=0$.

\subsection{Charge order, bond periodicity, and spin gap}

\begin{figure}
\centerline{\resizebox{3.2in}{!}{\includegraphics{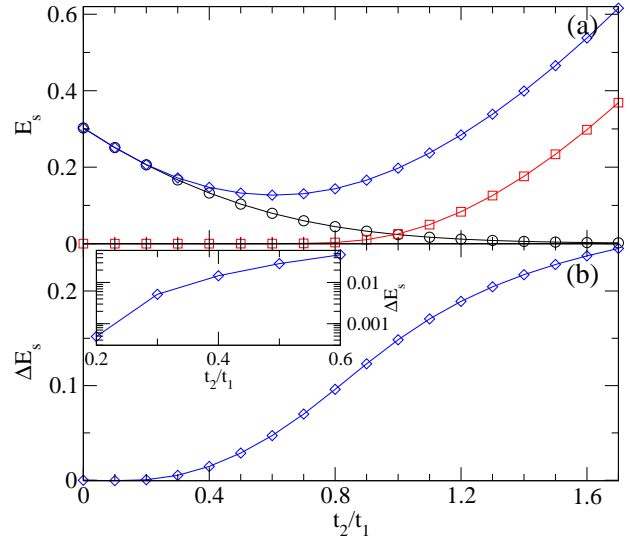}}}
\caption{(color online) (a) $E_s$, the per site spring constant energy
  term in Eq.~\ref{ham}, as a function of $t_2/t_1$ for a 256 site
  ladder with $U=V=0$.  Electron-phonon parameters are $\alpha_1=1.6$,
  $\alpha_2=0$ (circles); $\alpha_1=0$, $\alpha_2=1.6$ (squares); and
  $\alpha_1=1.6$, $\alpha_2=1.6$ (diamonds).  (b) Cooperative
  enhancement $\Delta E_s$ as a function of $t_2/t_1$ (see text). The
  inset shows the small $t_2/t_1$ region in more detail.}
\label{nonint}
\end{figure}
We first consider the solution of Eq.~\ref{ham} in the limit of zero
e-e interactions ($U=V=0$).  Because Eq.~\ref{ham} includes e-p
coupling in both the $t_1$ and $t_2$ directions, the variation of the
lattice distortion amplitude with the ratio $t_2/t_1$ is nontrivial.  The total
spring constant energy $E_s$, the sum of $K_1$ and $K_2$ terms in
Eq.~\ref{ham}, is a convenient measure of the strength of the lattice
distortion.  Fig.~\ref{nonint}(a) shows $E_s$ as a function of $t_2/t_1$
for a 256 site ladder.  Three different choices for e-p couplings are
shown: (i) $\alpha_1=\alpha$, $\alpha_2=0$; (ii) $\alpha_1=0$,
$\alpha_2=\alpha$; and (iii) $\alpha_1=\alpha_2=\alpha$. In the first
case with e-p coupling only along the zigzag direction, the lattice
distortion is strongest when $t_2=0$, and vanishes continuously at
$(t_2/t_1)_c$. In the second case with e-p coupling only along the
$t_2$ bonds, the strength of the distortion increases with
$t_2$. Putting these two effects together, the strength of the 
lattice distortion in
the general case with $\alpha_1$ and $\alpha_2$ nonzero shows a
minimum at an intermediate value of $t_2/t_1$, as seen in Fig.~\ref{nonint}.
 The precise $t_2/t_1$
value of this minimum will depend on the specific choices for
$\alpha_1$ and $\alpha_2$.

Fig.~\ref{nonint}(a) also shows that the $t_2$ and $t_1$ lattice
distortions act {\it cooperatively}--the total lattice distortion for
both $\alpha_1$ and $\alpha_2$ nonzero is considerably stronger than
the sum of the independent distortions.  In Fig.~\ref{nonint}(b), we
plot a quantitative measure of the cooperative enhancement, $\Delta
E_s=E_s(\alpha_1=\alpha_2=\alpha)
-(E_s(\alpha_1=\alpha,\alpha_2=0)+E_s(\alpha_1=0,\alpha_2=\alpha))$,
as a function of $t_2/t_1$.  $\Delta E_s$ increases continuously with
$t_2/t_1$, and as shown in the inset is nonzero even for small values
of $t_2/t_1$.  As we will discuss further in Section \ref{discussion},
this cooperative effect also has potentially important consequences
for materials---if both $\alpha_1$ and $\alpha_2$ are nonzero, near
$(t_2/t_1)_c$ the lattice distortion changes abruptly.

\begin{figure}
\centerline{\resizebox{3.2in}{!}{\includegraphics{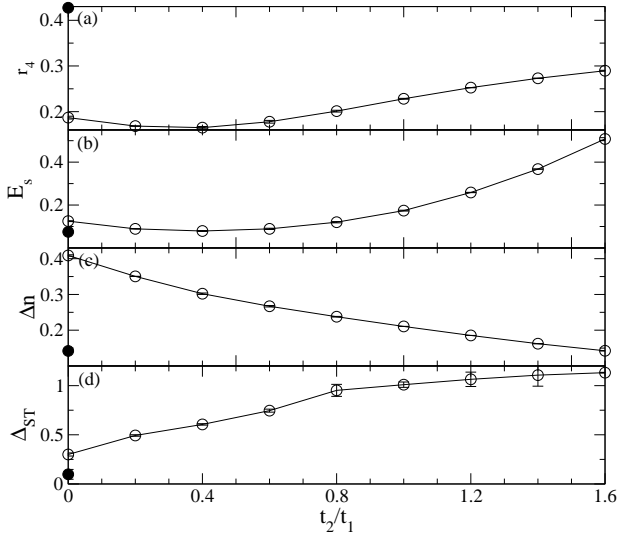}}}
\caption{Order parameters versus $t_2/t_1$ for the zigzag ladder with
  $U=6$, $V=1$, and $\alpha=1.6$.  Results are finite-size scaled from
  20, 28, 36, and 60 site lattices.  (a) 4k$_F$ component of the
  lattice distortion, and (b) spring energy per site, $E_s$, (c)
  charge disproportionation $\Delta$n, and (d) singlet-triplet gap
  $\Delta_{\rm{ST}}$.  Open symbols are for $V=V_1=V_2$, and filled
  symbols at $t_2/t_1=0$ are for $V=V_1$, $V_2=0$ (see text).}
\label{data}
\end{figure}
A key difference between the $\rho=\frac{1}{2}$ zigzag ladder and the
single chain is that in the ladder the bond pattern of the PEC
remains SMWM regardless of the strength of e-e interactions.
In 1D the displacement of site $j$ from equilibrium, $u_j$
($\Delta_{j,j+1}=u_{j+1}-u_j$ in Eq.~\ref{ham}),
can be written as \cite{Ung94a}
\begin{equation}
u_j=u_0[r_2\cos(2k_{\rm{F}}j-\theta_2) + r_4\cos(4k_{\rm{F}}j-\theta_4)],
\label{r2r4}
\end{equation}
where $r_2$ and $r_4$ are relative components of the period-4
2k$_{\rm{F}}$ and period-2 4k$_{\rm{F}}$ lattice distortions and $u_0$
is an overall amplitude. $r_2$ and $r_4$ are normalized such that
$r_2+r_4=1$.  The SMWM bond pattern corresponds to $r_4<0.41$ while
the SWSW$^\prime$ pattern\cite{Ung94a} has $r_4>0.41$. In both cases
$\theta_2=\frac{\pi}{4}$ and $\theta_4=0$.  Note that $\Delta_{j,j+2}$
in Eq.~\ref{ham} is an independent lattice distortion of the $t_2$
bonds and has no counterpart in the 1D model.

While the PEC occurs in the thermodynamic limit for
$t_2/t_1<(t_2/t_1)_c$ and $\alpha=0^+$, in finite lattices a finite
e-p coupling is required to observe the broken-symmetry ground state.
We have chosen the e-p coupling strengths $\alpha$ and $g$ just larger
than the minimum required for the ground and triplet states to be
Peierls distorted.  The strength of the lattice distortion depends on
the values of $\alpha$ and $\beta$ and therefore results here should
not be directly compared to experimental values. 

\begin{figure}
\centerline{\resizebox{3.2in}{!}{\includegraphics{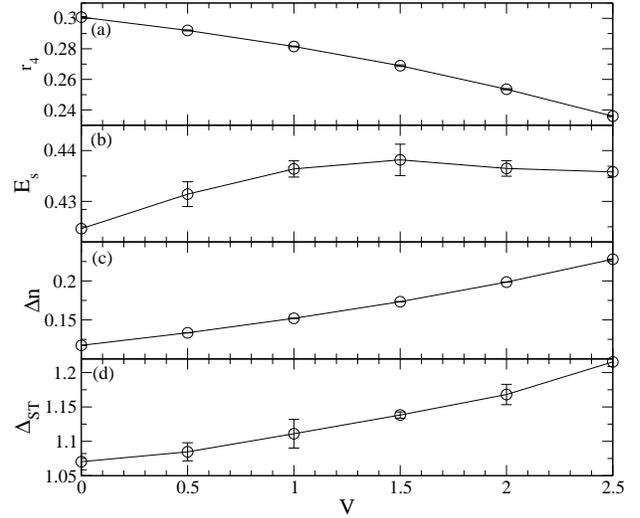}}}
\caption{Dependence of order parameters on $V$.  Results are
  finite-size scaled from 20, 28, 36, and 60 site lattices.
  Quantities plotted in panels (a)-(d) are as defined in
  Fig.~\ref{data}.  Here $t_2/t_1=1.5$, $U=6$, and $\alpha=1.6$. }
\label{vdata}
\end{figure}
In Fig.~\ref{data} we plot the finite-size scaled values of several
order parameters determined self-consistently versus $t_2/t_1$ for
$\alpha=1.6$, $U=6$, and $V=1$.  Exact and MPS-QMC calculations were
performed for 20, 28, 36, and 60 site lattices.  Details of the
finite-size extrapolation are given in Appendix \ref{appendixb}.

 Fig.~\ref{data}(a) shows the 4k$_F$ lattice
distortion strength and Fig.~\ref{data}(b) the spring constant
energy per site (including both $t_1$ and $t_2$ lattice distortions).
Fig.~\ref{data}(c) shows the charge disproportionation $\Delta n$,
defined as the difference between the charge density on charge-rich
and charge-poor sites.  The singlet-triplet gap defined as
$\Delta_{\rm{ST}}=E(S=1)-E(S=0)$ is plotted in Fig.~\ref{data}(d).
Note that two different `1D' limits are shown in Fig.~\ref{data} at
$t_2/t_1=0$: either including the second neighbor Coulomb interaction
($V_2=V$, open symbols), or with only nearest-neighbor Coulomb
interactions ($V_2=0$, filled symbols).

Focusing first on the $t_2/t_1=0$ limit, the bond pattern is SMWM in
the case where $V_2$ is nonzero, as is found in the 1D weakly
correlated band\cite{Ung94a}. In the more traditional 1D with no
second-neighbor $V_2$, the bond distortion pattern is SWSW$^\prime$
for $U=6$, $V=1$ with $r_4>0.41$ (filled circle). For equal e-p
coupling, the lattice distortion energy, $\Delta n$, and
$\Delta_{\rm{ST}}$ are all stronger when the bond pattern is SMWM.
Away from the 1D limit, $r_4$ increases with increasing $t_2/t_1$ but
is always less than the 0.41 that would be necessary to reach the
SWSW$^\prime$ bond pattern.  The amplitude of the lattice distortion
measured by the spring constant energy in Fig.~\ref{data}(b) shows
a minima at intermediate $t_2/t_1$ as in the non-interacting case.
Surprisingly, $\Delta n$ {\it decreases} as the lattice distortion
becomes stronger--this is one significant difference from the 1D chain
BCDW state where $\Delta n$ follows the strength of the bond
distortion.  Reference \cite{Clay05a} showed that the spin gap
in the PEC state of the zigzag ladder is larger than the gap in the
single chain having the same $\Delta n$.  As Fig.~\ref{data}(d)
shows, the gap in the ladder increases with $t_2/t_1$, and is largest
near $(t_2/t_1)_c$.

The $U$ interaction weakens the PEC in the zigzag ladder (not
shown). This decrease is, however, weaker than in the 1D
chain\cite{Clay05a}. In Fig.~\ref{vdata} we show the results of
varying $V$ while keeping other parameters ($t_2/t_1=1.5$, $U=6$, and
$\alpha=1.6$) fixed.  Even at large $V$ ($V>U/2$), as expected we did
not find any transition to a distinct WC state; in all cases the CO
pattern is $\cdots1100\cdots$ along the zigzag direction.  Contrary to
what occurs in the 1D limit, the 4k$_F$ component of the bond
distortion actually {\it decreases} with increasing $V$.  While in 1D
$V$ destabilizes the $\cdots1100\cdots$ CO, in the zigzag ladder, this
effect in the ladder along the $t_1$ direction is countered by $V_2$,
which prefers $\cdots1010\cdots$ order along the $t_2$ directions. A
similar effect is found in the $\frac{1}{4}$-filled PEC state in the
2D anisotropic triangular lattice---there also $V$ strengthens the
$\cdots1100\cdots$ CO provided $V$ is not too large \cite{Dayal11a}.
The most 
\begin{figure*}[t]
\centerline{\resizebox{5.5in}{!}{\includegraphics{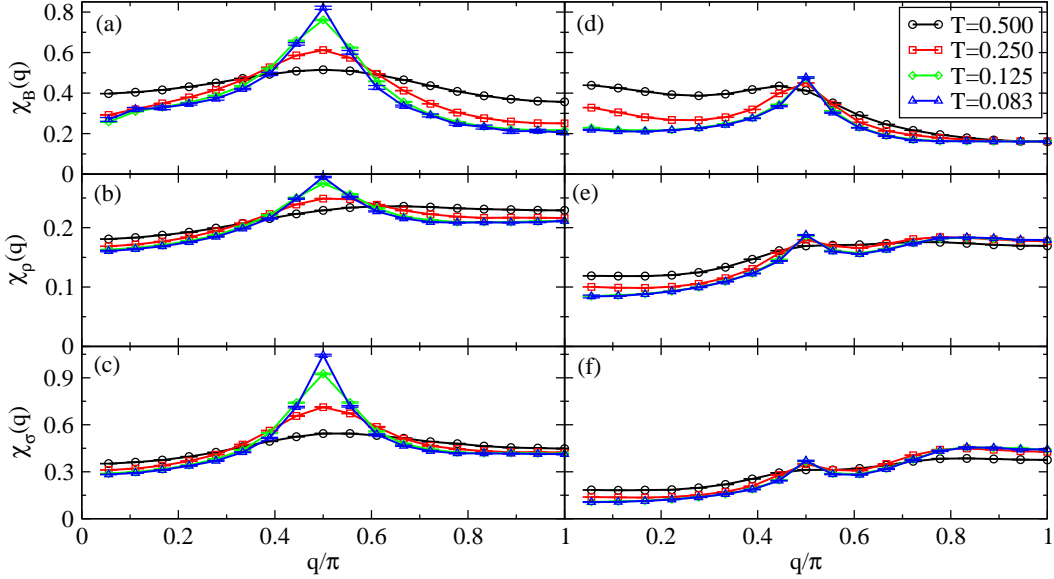}}}
\caption{(color online) Representative temperature-dependent QMC
  results for bond ((a) and (d)), charge ((b) and (e)), and spin
  susceptibilities ((c) and (f)) for a 36 site zigzag ladder with
  $U=4$, $V=0$, and $\alpha=0$. $t_2/t_1$ is 0.4 in panels (a)-(c),
  and 1.4 in panels (d)-(f).}
\label{qmcdata1}
\end{figure*}
striking result of Fig.~\ref{vdata} is the very large spin
gap that is obtained when both $t_2/t_1$ and $V$ are moderately
large. We will return to this point later in Section \ref{discussion}.

\subsection{Thermodynamics and spinon binding}

To understand the thermodynamics of the zigzag ladder we present our
results from complementary calculations of (i)
wavenumber-dependent susceptibilities, and (ii) the nature of higher
spin states.  Finite-temperature calculations of susceptibilities here
are done within the static undistorted lattice.

\subsubsection{Temperature dependence of susceptibilities}

We calculate wavenumber dependent charge ($\chi_\rho(q)$), spin
($\chi_\sigma(q)$), and bond susceptibilities ($\chi_B(q)$), defined
as
\begin{equation}
  \chi_x(q) = \frac{1}{N} \sum_{j,k} e^{iq(j-k)} \int_{0}^{\beta} d\tau
 \langle O^x_j(\tau)O^x_k(0)\rangle.
\label{eqn:susceptibility}
\end{equation}
In Eq.~\ref{eqn:susceptibility}, $O^\rho_j = n_{j,\uparrow} +
n_{j,\downarrow}$, $O^\sigma_j = n_{j,\uparrow} - n_{j,\downarrow}$,
and $O^B_j = B_{j,j+1}$, for charge, spin, and bond order
susceptibilities, respectively.  $\beta$ is the inverse temperature in
units of $t_1$. To facilitate comparison with 1D, the $q$ is again
taken to be one dimensional as in Eq.~\ref{bands}. The presence and
periodicity of charge and bond order can be determined from
divergences of the charge or bond-order susceptibility as
$\beta\rightarrow\infty$ Within determinantal QMC methods finite $V$
leads to a significantly worse Fermion sign problem and hence we
present results only for nonzero $U$ but $V=0$.  However, as shown in
Fig.~\ref{vdata}(a), $V$ does not change the pattern of the bond
distortion.  We therefore expect that our results here are
representative for arbitrary $U$ and $V$.

Fig.~\ref{qmcdata1} shows the bond, charge, and spin susceptibilities
as a function of $q$ and temperature for a $N=36$ site ladder with
$U=4$, $V=0$, and two representative values of $t_2/t_1$, 0.4
(Fig.~\ref{qmcdata1}(a)-(c)) and 1.4 (Fig.~\ref{qmcdata1}(d)-(f)).  At
low temperature, the bond, charge, and spin susceptibilities all peak
at 2k$_F$, consistent with period-four order of charges and bonds as
expected.  The 2k$_F$ peak in the spin susceptibility corresponds to
the short-range AFM spin order found in the 1D $\frac{1}{4}$-filled
chain.  As seen in Fig.~\ref{qmcdata1}(c) and (f), the spin
susceptibility converges to a finite value as $q\rightarrow 0$ at low
temperatures.  This is consistent with the expectation that no SG
exists in the ladder in the limit of zero e-p coupling.  Several
differences can be noted when comparing results for small and large
$t_2/t_1$. First, the 2$k_F$ bond and charge response as T$\rightarrow
0$ is stronger for small $t_2/t_1$. This reflects the larger amplitude
distortion ($\Delta n$ and bond order) found in the 1D
weakly-correlated limit (see Fig.~\ref{data}(b)-(c)).  The 2$k_F$ spin
response becomes weaker with increasing $t_2/t_1$, and
$\chi_\sigma(q)$ is reduced for small $q$.  These changes reflect the
increasing strength of the spin-singlet bond along the $t_1$ direction
with increasing $t_2/t_1$, which in turn leads to the increase in
$\Delta_{\rm{ST}}$.  For larger $t_2/t_1$, a broad plateau appears at
$q\approx 3\pi/4$ in the charge and spin susceptibilities. This
plateau is however non-divergent---while it is significant at high and
intermediate T, at low temperatures the 2$k_F$ response becomes
stronger. We will show below that this plateau reflects the binding of
spinon excitations.

Summarizing, the above numerical results show that in the zigzag
ladder, no separate high-temperature ordering is expected; instead the
ladder is metallic at high temperature and as temperature decreases a
single transition to the PEC state with SG takes place. This is in
contrast to what happens in 1D \cite{Clay07a}.

\subsubsection{Spinon excitations}

\begin{figure}
\centerline{\resizebox{3.4in}{!}{\includegraphics{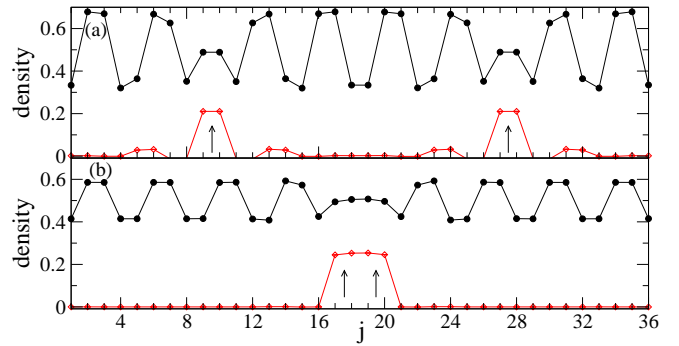}}}
\caption{(color online) MPS-QMC charge and spin density in the triplet
  state versus site index $j$ along the $t_1$ direction.  Parameters
  are $N=36$, $U=6$, $V=1$, and $\alpha=1.6$.  In panels (a) and (b),
  $t_2/t_1$=0.2 and 1.4, respectively.  Circles (diamonds) are charge
  (spin) density. Arrows indicate location of spinon defects (see
  text). Statistical errors are smaller than symbol sizes.}
\label{mps36site}
\end{figure}

The structure of excitations out of the ground state provides an
alternate way to understand the thermodynamics of a
strongly-correlated system.  In the 1D limit, flipping one spin in the
PEC results in two spinons which are unbound and hence separated by a
distance of $N/2$ sites on the lattice.  We have performed
self-consistent calculations within Eq.~\ref{ham} of excited spin
triplet ($S=1$) excitations to detect and study such spinon
excitations in the zigzag ladder.  Our calculations are for moderately
large e-p interactions so that the widths of the charge-spin solitons
are relatively narrow. This is necessary in order to prevent overlaps
between the solitons.  In Fig.~\ref{mps36site} we show the charge and
spin densities in the $S=1$ state of a 36 site zigzag ladder for small
and large $t_2/t_1$.  Spinon excitations are identified as defects in
the $\cdots1100\cdots$ charge order, and also from their large local
spin densities \cite{Clay07a}.

For small $t_2/t_1$ (Fig.~\ref{mps36site}(a)), flipping a single spin
results in two separated spinons as in the 1D chain.  The repulsive
interaction between spinons results in their separating to opposite
positions on the periodic lattice.  As indicated in the figure, each
spinon occupies two lattice sites, and there occur both charge and
spin modulations.  The charge density at each spinon site is 0.5; the
spin densities are also equal on the two sites.  The PEC charge and
bond distortions on either side of a spinon are out of phase with each
other by two lattice sites.  The charge densities on the two sites at
the center of the are 0.5, 0.5 (see Fig.~\ref{mps36site}(a)).  With
increasing $t_2/t_1$, the lattice separation between the spinons
decreases, indicating binding. For the e-p coupling of
Fig.~\ref{mps36site}, at approximately $t_2/t_2\approx 1.0$ the bound
spinons form a single excitation that occupies 4 lattice sites with
approximately uniform charge density of 0.5 on each site
(Fig.~\ref{mps36site}(b)).  For spin states higher than 1 (not shown
here), we found that spinons are bound in pairs; for example in the
$S=2$ state there are two of the defects shown in
Fig.~\ref{mps36site}(b), separated by the maximum possible lattice
spacing. Spinon binding is usually associated with an increase in the
singlet-triplet gap, and can thus explain the observation
the(Fig.~\ref{data}(d)) that for fixed $V$, $\Delta_{\rm{ST}}$
increases with increasing $t_2/t_1$, even though $\Delta n$ decreases
at the same time.

The thermodynamic behavior is understood only when the complementary
calculations of Fig.~\ref{qmcdata1} and \ref{mps36site} are taken
together. Fig.~\ref{mps36site} shows spinon creation upon a single
spin excitation, while the thermodynamic properties at intermediate
and high temperatures are dominated by high-spin configurations
containing multiple spinons. Thus the signatures of spinons, including
spinon binding can also are to be be found in the finite-temperature
QMC results of Fig.~\ref{qmcdata1}.  In the 1D limit spinons have
local 4k$_F$ charge or bond order \cite{Clay07a}.  As in the 1D chain
(reference \cite{Clay07a}), when $t_2/t_1$ is small we find that the
4k$_F$ bond and charge susceptibilities {\it increase} with
temperature (see Fig.~\ref{qmcdata1}(a)-(b) at $q=\pi$).

When $t_2/t_1$ is of order unity, several differences as seen in the
susceptibilities that can be correlated with the binding of spinons
shown in Fig.~\ref{mps36site}.  First, at intermediate temperatures,
the charge and spin susceptibilities have a broad plateau at $q\approx
3\pi/4$. When spinons bind, the larger size of these defects moves the
charge and spin response away from 4k$_F$ and towards a smaller
wavenumber.  Second, in this parameter region the 4k$_F$ bond
susceptibility remains small regardless of temperature and instead at
higher temperatures a broad response from $0<q<\pi/2$ appears
(Fig.~\ref{qmcdata1}(d)).  This small-$q$ (long wavelength) bond order
response is a result of increasing numbers of bound spinons created in
high-spin configurations.  Each high-spin configuration has multiple
bound spinons equally separated from each other, giving a
long-wavelength bond order distortion with $0<q<\pi/2$.  Related to
this is the observation that $\chi_B(2k_F)$ varies only weakly with
temperature in the ladder limit.  In the 1D limit, each spinon
separates regions where the PEC is out of phase by two lattice units
(Fig.~\ref{mps36site}(a)); due to this phase difference the
introduction of spinons will lead to a rapid decrease in the 2k$_F$
response. On the other hand in the ladder limit, the PEC remains ``in
phase'' on either side of bound spinons (Fig.~\ref{mps36site}(b)),
resulting in a weaker temperature dependence of the 2k$_F$ bond
susceptibility---bound spinons suppress the BCDW locally, but do not
disturb the overall phase of the density wave as do single spinons.

\section{Discussion}
\label{discussion}

We have shown that the $\frac{1}{4}$-filled correlated zigzag ladder
has the following properties: First, due to the geometry of this
lattice, the WC state driven by strong nearest-neighbor Coulomb
interactions in 1D is strongly suppressed. Instead, a single PEC state
with CO pattern $\cdots1100\cdots$ occurs over a wide range of
$t_2/t_1$ ($0 \leq t_2/t_1 \leq 1.707$).  Second, unlike in the 1D
$\frac{1}{4}$-filled band the bond order pattern in the zigzag ladder
is $SMWM$ for all Coulomb interactions, and the bond distortion
$SWSW^\prime$ never occurs.  Third, with increasing $t_2/t_1$ from the
1D limit, the CO amplitude decreases for fixed $V$ until it vanishes
above a critical value. Surprisingly, the singlet-triplet gap
increases in magnitude at the same time. We have shown that this
increase is accompanied by the binding of spinon excitations. The
singlet-triplet gap can be very large when $t_2/t_1$ and $V$ are both
moderately large (see Fig.~\ref{vdata}(d)).  Finally, our QMC
calculations show that in the zigzag ladder there is no distinct
intermediate state between the low temperature PEC and the high
temperature metallic state, and there is a single metal-insulator
transition that is accompanied by simultaneous bond and charge
distortions and spin gap.  The existence of simultaneous transitions
is expected here as the spin-singlet here is of the ``intra-dimer''
type as shown in Fig.\ref{1d}(c)-(d).

In the isotropic $\frac{1}{4}$-filled band 2D square lattice, adding a
frustrating diagonal $t^\prime$ bond results in a transition from a
uniform state to a paired PEC state once $t^\prime$ exceeds a critical
value \cite{Li10a,Dayal11a}. As with spin systems, the zigzag ladder
provides a simple model for the study of the frustration-driven PEC
state in 2D, with the difference that the frustration does not create
the SG state in the ladder, but rather enhances the SG that is already
present in the unfrustrated model.  
This enhancement can be large due to the cooperative interaction between 
the two kinds of electron-phonon interactions that are possible in the 
ladder, as shown in Fig.~\ref{nonint}.
We discuss below the implications
of our work for understanding experiments on $\frac{1}{4}$-filled
materials in general (including both 2D and 3D). We first consider
specifically those materials which have been suggested to be ladders
based on their crystal structures.  We then also consider the broader
class of $\frac{1}{4}$-filled band materials. Ideally, this second
class of materials require understanding of the excitations and
thermodynamics of the higher dimensional PEC, which is a much more
formidable task than the ladder calculations. We nevertheless point
out that broad conclusions can be drawn for these systems based on our
current work.  The majority of the materials we consider belong to the
large family of low dimensional organic charge transfer solids (CTS),
within which many examples of SG ground states are found with quasi-1D
and 2D lattice \cite{Ishiguro}.  We also point out a few inorganic
$\frac{1}{4}$-filled materials that exhibit similar transitions.

\subsection{$\frac{1}{4}$-filled ladder candidates}

While most ladder materials found have been $\frac{1}{2}$-filled
\cite{Dagotto96a}, the family of CTS materials (DTTTF)$_2$M(mnt)$_2$,
where M is a metal ion, are likely candidates for $\frac{1}{4}$-filled
band ladders \cite{Rovira00a}.  In this series of compounds, the M=Au
and M=Cu (for both the metal ion is diamagnetic) have been studied in
most detail
\cite{Rovira00a,Ribera99a,Wesolowski03a,Ribas05a,Musfeldt08a}.
Structurally, these materials consist of pairs of DTTTF stacks, each
with $\frac{1}{4}$-filled band of holes, separated by stacks of
M(mnt)$_2$ which are Mott-Hubbard semiconductors with
$\frac{1}{2}$-filled electron bands.  It is then likely that the each
pair of DTTTF stacks behaves as ladders. Note that the stack direction
corresponds to the $t_2$ direction within our model, and hence these
systems lie in the parameter regime $t_2>t_1$. The very
large\cite{Ribas05a} T$_{\rm{SG}}$ in (DTTTF)$_2$Au(mnt)$_2$ ($\sim$70
K) and (DTTTF)$_2$Cu(mnt)$_2$ ($\sim$90K), which are nearly an order
of magnitude larger than the spin-Peierls transition temperatures in
the 1D $\frac{1}{4}$-filled systems \cite{Pouget88a}, supports the
conjecture that these systems are ladders.
  
The MI and SG transitions are, however, distinct in these compounds,
which would argue against the zigzag ladder picture, at least its
simplest version. For M=Au, a broad MI transition occurs at
T$_{\rm{MI}}\approx 220$ K, followed by a decrease in the magnetic
susceptibility at 70 K \cite{Ribera99a}.  Below the MI transition,
diffuse X-ray scattering at $b^\star/2$ indicates dimerization along
$t_2$, but broad line widths suggest the dimerization order is not
long-ranged \cite{Ribera99a}.  The M=Cu salt is isostructural to the
Au salt with slightly smaller lattice parameters due to the smaller
metal ion. The MI transition for M=Cu occurs at 235 K, and unlike the
Au salt is a sharp, second-order phase transition
that is accompanied by doubling of the unit cell in the ladder 
direction \cite{Ribas05a}. Changes in the optical properties at T$_{\rm{MI}}$
and T$_{\rm{SG}}$ for the two salts are also different
\cite{Wesolowski03a,Musfeldt08a}. For M=Au, at T$_{\rm{MI}}$ symmetry
breaking occurs along the rung direction (perpendicular to the DTTTF
stacks), while at T$_{\rm{SG}}$, symmetry breaking is predominantly
along the stacks \cite{Wesolowski03a}.  In contrast, optical response
indicates symmetry breaking in both rung and stack directions at
T$_{\rm{MI}}$ for M=Cu \cite{Musfeldt08a}.

We believe that the $\frac{1}{4}$-filled band zigzag ladder model is
nevertheless a valid description for both M = Au and Cu at low
temperatures. The only other competing model for these systems is the
rectangular ladder model \cite{Wesolowski03a}, wherein each DTTTF
molecule is coupled to a single other such molecule on the neighboring
stack. Such a description would be against the known crystal
structures \cite{Rovira00a}. Furthermore, within the rectangular
ladder model, there needs to occur a high temperature metal-insulator
transition accompanied by in-phase bond dimerization, such that each
dimer of DTTTF molecules has a single electron; the ladder after
dimerization would be akin to rectangular spin ladder, which has SG
for all interstack spin exchange.  The two stacks need to be identical
within the model and hence there is no symmetry breaking within the
rectangular ladder scenario along the rung direction at any
temperature. Neither is there any CO within the model, in
contradiction to what is found in optical measurements
\cite{Musfeldt08a}.

There can be several different reasons why the high temperature MI
transition occurs within the zigzag ladder scenario. First, muon spin
rotation experiments suggest significant interladder coupling
\cite{Arcon99a}, that has been ignored in our isolated ladder
model. Second, our model does not take into account the
temperature-dependent lattice expansion that is common to CTS
crystals.  Note that lattice expansion will affect the interstack
hopping $t_1$ much more strongly than the intrastack hopping $t_2$. It
is then conceivable that at high temperatures the interstack distance
is large enough (and $t_1$ is small enough) that $t_2/t_1$ is greater
than the critical value 1.707 and the systems behave as independent
chains.  The lattice contracts at reduced temperatures, increasing
$t_1$ and reducing $t_2/t_1$, when the systems exhibit zigzag ladder
behavior. This would require $t_2/t_1$ close to 1.7 in the
experimental systems, which is indeed close to ratio of the calculated
hopping integrals for the M = Au system \cite{Ribera99a}.  As seen in
Fig.~\ref{data}(d), large $t_2/t_1$ would be in agreement with the
unusually large SG seen in the (DTTTF)$_2$M(mnt)$_2$. There are
however additional complications. During the synthesis of the
(DTTTF)$_2$M(mnt)$_2$ salt, the 1:1 salt (DTTTF)M(mnt)$_2$ is also
produced and crystals of the 1:2 salt must therefore be separated from
this mixture for experiments \cite{Ribas05a}. Relative to other CTS,
available experimental data is thus more limited.  Experimental
determinations of the pattern of the CO below T$_{\rm{SG}}$ (and
above, if any), and of the temperature-dependent lattice distortion
are necessary for resolution of the above issues.

\subsection{General classification of $\frac{1}{4}$-filled materials}

As discussed in Section \ref{1dsummary}, in 1D the SG and MI
transitions are distinct when the ground state broken-symmetry state
has the bond pattern SWSW$^\prime$, but are coupled together in a
single transition when the bond pattern is SMWM as occurs in the
zigzag ladder we have considered here. In 1D, the strength of e-e
interactions determines which bond pattern is favored. As SG
transitions are found in a number of $\frac{1}{4}$-filled materials
with 2D as well as 3D lattices, generalizations of these results to
higher lattice dimensionality are of great interest. Our results in
Section \ref{results} for the zigzag ladder show that in dimensions
greater than one, e-e interaction strength does {\it not necessarily}
determine the bond distortion pattern and thermodynamics---lattice
structure and frustration are also important.

We point out an empirical criterion here that rationalizes separate
versus coupled SG--MI transitions in $\frac{1}{4}$-filled band
materials at large, that we arrive at by simply extrapolating from the
1D and zigzag ladder results.  Our observation is that if the low
temperature structure is such that the singlet bond is interdimer, and
the strongest bond is between intradimer charge-rich and charge-poor
sites (Fig.~\ref{1d}(b)), there occur distinct transitions involving charge
and spin degrees of freedom.  Conversely, if the intradimer bond is
between a pair of charge-rich sites (Fig.~\ref{1d}(c)), and the SG is due to
intradimer spin-singlets, there is a single coupled SG-MI transition.
In this latter case in general T$_{SG}$ is high.  The first of these
two observations was noted previously by Mori \cite{Mori99b}.  We do
not have any microscopic calculation to justify these conclusions for
2D and 3D; they are based on the mappings of Fig.~\ref{1d}. In
Table~\ref{materialstab} we give a list of materials for which the
bond and/or charge ordering pattern below T$_{SG}$ is known. In all
cases our simple criterion appears to be valid (see below).  Two of
the entries require additional explanation.  (TMTTF)$_2$PF$_6$ is
already insulating at room temperature because of lattice
dimerization; the high temperature CO at 70 K here is of the WC type,
but there is strong redistribution of the charge below T$_{SG}$
\cite{Nakamura07a}, clearly placing this system into the category of
materials that undergoes two transitions \cite{Clay07a}. The low
T$_{SG}$ is also a signature of this. The situation with
$\alpha^\prime$-NaV$_2$O$_5$ is exactly the opposite. This system is
also insulating already at high temperature, but now the
charge-ordering and spin-gap transitions occur at the same
temperature, indicating that these transitions are coupled.  While
Table~\ref{materialstab} is meant to be representative and not
comprehensive, we are unaware of examples where our criterion
fails. We describe individual materials in detail below.
\begin{table}
\begin{tabular}{l|l|D{.}{.}{0}|D{.}{.}{0}|l}
\hline
Material & D-n & \multicolumn{1}{l|}{T$_{\rm{SG}}$ } & 
\multicolumn{1}{l|}{T$_{\rm{CB}}$}  & Ref.\\
 & & (K) & (K) & \\
\hline
MEM(TCNQ)$_2$ & 1D-2 & 17 & 335  & 42,43 \\%\cite{Huizinga79a,Visser83a} \\
(TMTTF)$_2$PF$_6$ & 1D-2 & 19 & 70 & 38,44 \\% \cite{Pouget88a,Nad06a} \\
$\theta$-(ET)$_2$RbZn(SCN)$_4$ & 2D-2 & 20 & 195 & 45-48 \\%\cite{Mori98b,Miyagawa00a,Watanabe04a,Watanabe07a} \\
EtMe$_3$P[Pd(dmit)$_2$]$_2$ & 2D-2  & 25 & >300 & 49,50 \\%\cite{Kato06a,Tamura06a} \\
$\beta^{\prime\prime}$-DODHT)$_2$PF$_6$ & 2D-2 & 40 & 255  & 51,52 \\% \cite{Nishikawa02a,Nishikawa05a}  \\
(DMe-DCNQI)$_2$Ag & 1D-2 & 80 & 100 & 53 \\% \cite{Werner88a} \\
$\alpha$-NaV$_2$O$_5$ & 2D-1 & 34 & 34 & 54 \\%\cite{Isobe96a}\\
$\alpha^\prime$-(ET)$_2$I$_3$ & 2D-1 & 135 & 135 & 55,56 \\% \cite{Rothaemel86a,Kakiuchi07b} \\
(BDTFP)$_2$PF$_6$(PhCl)$_{0.5}$& 1D-1 &175 & 175 & 57-59 \\%\cite{Ise01a,Uruichi02a,Uruichi03a} \\
CuIr$_2$S$_4$ & 3D-1 & 230 & 230 & 60,61 \\%\cite{Radaelli02a,Khomskii05a} \\
(EDO-TTF)$_2$PF$_6$  & 1D-1 & 280 & 280 & 62,63 \\% \cite{Ota02a,Drozdova04a} \\
\hline
\end{tabular}
\caption{$\frac{1}{4}$-filled materials with spin gapped ground
  states. For each the dimensionality and number of transitions D-n,
  spin-gap transition temperature T$_{\rm{SG}}$, charge--bond ordering
  temperature T$_{\rm{CB}}$, and references are listed.}
\label{materialstab}
\end{table}

\subsubsection{Two transitions: inter-dimer singlet formation}

We have previously discussed quasi-1D CTS where two transitions occur
\cite{Clay07a}.  MEM(TCNQ)$_2$ is one example where the low
temperature bond pattern has been well characterized by neutron
scattering \cite{Visser83a}.  In MEM(TCNQ)$_2$, the MI transition
occurs near room temperature at T$_{\rm{MI}}$=335 K, followed by a SG
transition at T$_{\rm{SG}}$=17.4 K \cite{Huizinga79a,Visser83a}. The
bond pattern at low temperature is SWSW$^\prime$ \cite{Visser83a}.
The (TMTTF)$_2$X series is another quasi-1D example where CO and SG
occur at different temperatures \cite{Chow00a,Foury-Leylekian04a,Nad06a,Fujiyama06a,Clay07a}.

In the 2D $\theta$-(BEDT-TTF)$_2$MM$^\prime$(SCN)$_4$ series, CO
occurs at the high temperature MI transition
\cite{Mori98b,Miyagawa00a} followed by the SG transition at
T$_{\rm{SG}}\sim$20 K \cite{Mori98b,Watanabe04a,Watanabe07a}. X-ray
studies in the temperature range T$_{\rm{SG}}<$T$<$T$_{\rm{CO}}$ show
that the strongest bond orders in the 2D BEDT-TTF layers are between
molecules with large and small charge density (see Fig.7 in reference
\cite{Watanabe04a}).  These and lower temperature X-ray
results\cite{Watanabe07a} have indicated that the spin-singlets in the
SG phase are located on the inter-dimer bonds, as in the SWSW$^\prime$
bond pattern in 1D.

Yet another 2D material that very clearly shows two transitions and in
which the charge-bond distortion patterns are known is
EtMe$_3$P[Pd(dmit)$_2$]$_2$ \cite{Kato06a,Tamura06a}. The material is
semiconducting already at 300 K (T$_{MI} > 300$ K). The magnetic
susceptibility at high temperatures corresponds to that of a
Heisenberg antiferromagnet on a triangular lattice with $J=250$
K. Below a relatively low T$_{SG}$=25 K the system enters a distorted
phase, with the intermolecular bond distortion pattern clearly of the
dimerized dimer type, and the strongest bonds between charge-rich and
charge-poor sites \cite{Tamura06a}.  This material undergoes
superconducting transition under pressure \cite{Shimizu07a}.

\subsubsection{Coupled transitions: intra-dimer singlet formation}

In cases where T$_{\rm{MI}}$ and T$_{\rm{SG}}$ coincide, the spin gap
transition temperature tends to be quite high. Examples here include
(EDO-TTF)$_2$X, which shows a first order MI transition at high
temperature, 280 K for X=PF$_6$ and 268 K for X=AsF$_6$
\cite{Ota02a}. This transition coincides with T$_{\rm{SG}}$
\cite{Ota02a}. Optical experiments determined that the charge order
pattern in the low temperature phase is $\cdots$1100$\cdots$, with the
strongest bond between molecules with large charge density
\cite{Drozdova04a}.  A coupled SG--MI transition occurs at
T$_{\rm{MI}}$=175 K in (BDTFP)$_2$PF$_6$(PhCl)$_{0.5}$
\cite{Ise01a}. While structurally (BDTFP)$_2$PF$_6$(PhCl)$_{0.5}$
appears to be ladder-like \cite{Ise01a,Clay05a}, X-ray
\cite{Uruichi02a} and optical measurements \cite{Uruichi03a} show that
in the low-temperature phase tetramerization takes place along the
stacks with the SMWM bond pattern \cite{Uruichi02a}. The coupled
SG--MI transitions found in these two materials are consistent with
our criterion above.

Beyond quasi-1D materials, in the 2D CTS examples can be found with
coupled transitions, which also typically take place at a relatively
high temperature.  In $\alpha$-(BEDT-TTF)$_2$I$_3$ T$_{\rm{SG}}$=135 K
coinciding with the MI transition \cite{Rothaemel86a}. Similar to
(EDO-TTF)$_2$X, the SG--MI transition is first order and coincides
with a large structural change \cite{Kakiuchi07b}. In the low
temperature phase, the strongest bond is again between the sites of
largest charge density \cite{Kakiuchi07b}.

Similar transitions are observed in inorganic $\frac{1}{4}$-filled
materials as well. The inorganic spinel CuIr$_2$S$_4$ is one example
in which the Ir-ions form the active sites with $\frac{3}{4}$-filled
electron band ($\frac{1}{4}$-filled hole band) \cite{Khomskii05a}. In
CuIr$_2$S$_4$ a coupled SG--MI transition occurs at 230 K, below which
the criss-cross chains of Ir-ions are charge-ordered as
Ir$^{4+}$-Ir$^{4+}$-Ir$^{3+}$-Ir$^{3+}$, with the strongest bonds
between the spin $\frac{1}{2}$ hole-rich Ir$^{4+}$ ions and weakest
bonds between the spin 0 hole-poor Ir$^{3+}$ ions.
\cite{Radaelli02a,Khomskii05a}. A more complex coupled transition
occurs in $\alpha^\prime$-NaV$_2$O$_5$, where a coupled CO-SG
transition occurs at 34 K within an insulating phase. Structurally
$\alpha^\prime$-NaV$_2$O$_5$ consists of rectangular V-ion based
ladders linked by zigzag V-V bonds. Below the transition the V-ions
are charge disproportionated and there occurs a period-4
V$^{4+}$-V$^{4+}$-V$^{5+}$-V$^{5+}$ CO within the zigzag links between
the rectangular ladders.  Once again, the strongest bonds are between
the spin $\frac{1}{2}$ electron-rich V$^{4+}$ ion pairs and the
weakest bonds between the spin 0 electron-poor V$^{5+}$ ion pairs
\cite{Mostovoy99a,Edegger06a}, in agreement with our criterion for
coupled CO-SG transition.

\subsection{Possible relationship with superconductivity}

We have recently suggested that superconductivity in
strongly-correlated $\frac{1}{4}$-filled systems is due to a
transition from an insulating PEC state to a paired-electron liquid
\cite{Mazumdar08a,Li10a,Dayal11a}. Within this model, the
spin-singlets of the PEC become mobile with further increase in
frustration. The fundamental theoretical picture is then analogous to
bipolaron theories of superconductivity \cite{Alexandrov94a}, with two
differences: (i) the pairing in our model is driven by
antiferromagnetic correlations (as opposed to very strong
e-p interactions that screen out the short-range Coulomb
repulsion), and (ii) nearly all the carriers are involved in the
pairing. The effective mass of the spin-bonded pairs is an important
parameter, and overly strong binding will reduce pair mobility.  This
would suggest that superconductivity is {\it more likely} in materials
with inter-dimer singlets. As noted in the above, such systems tend to
have the dimerized dimer structure. In contrast, in those materials
with intra-dimer singlets that form at high temperature, the stronger
pair binding would lead to a pair mobility too low to achieve
superconductivity--in these cases the ground state would remain an
insulating spin-gapped PEC with charge and bond order. It is
interesting to note that that such a correlation was suggested from
empirical observations alone by Mori \cite{Mori99b}.

\section{Acknowledgments}

This work was supported by the US Department of Energy grant
DE-FG02-06ER46315.  RTC thanks A. Sandvik for helpful discussions
regarding the MPS-QMC method. RTC thanks the University of Arizona,
the Boston University Condensed Matter Theory Visitor's program, and
the Institute for Solid State Physics of the University of Tokyo for
hospitality while on sabbatical.

\appendix

\section{MPS-QMC method}
\label{appendixa}

Matrix-product states are best known as the class of wavefunctions
ultimately reached by the DMRG process \cite{Ostlund95a} and provide
highly efficient representations of wavefunctions for quasi-1D quantum
systems. An alternate set of methods to DMRG has recently emerged
which combines a MPS wavefunction representation with MC sampling
\cite{Sandvik07a,Schuch08a}.  In this approach, rather than the
renormalization procedure used by DMRG, Monte Carlo (MC) sampling is
used to evaluate expectation values of the MPS wavefunction.  The use
of MC has certain advantages, such as potentially better computational
scaling\cite{Sandvik07a}, and the ease with which the method can be
parallelized through trivial parallelization of the MC averaging.
\begin{figure}
\centerline{\resizebox{3.2in}{!}{\includegraphics{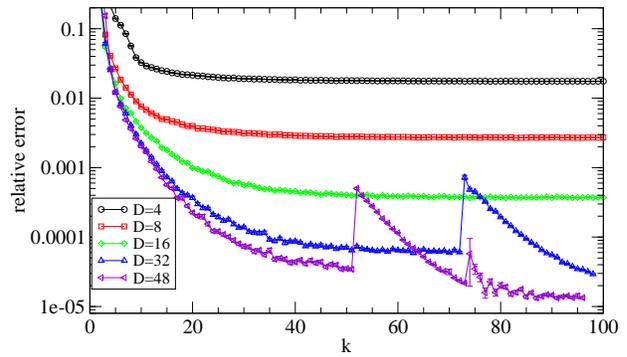}}}
\caption{(color online) Relative error in the ground state energy
  using MPS-QMC as a function of number of QMC steps $k$ and matrix
  size $D$ for a 20 site ladder with periodic boundary conditions and
  $t_1=t_2=1$, $U=6$, $V=1$, and $\alpha=0$.  Calculations for $D$=4,
  8, and 16 were single optimization runs, while those for 32 and 48
  were restarted after 50--75 steps (see text).}
\label{mps-test1}
\end{figure}

The MPS-QMC method we use is described in Reference
\cite{Sandvik07a}, where it was applied to the 1D quantum Ising
model in a transverse field. Here we show that the method can
successfully be applied to more complicated Fermion Hamiltonians.  In
the MPS representation, the wavefunction for a $N$ site system is
written as
\begin{equation}
|\Psi\rangle=\sum_{\{S\}}\rm{Tr}[A_1(s_1)A_2(s_2)\cdots A_N(s_N)]|s_1 s_2 \cdots s_N \rangle.
\label{mps}
\end{equation}
In Eq.~\ref{mps}, $|S\rangle=|s_1 s_2 \cdots s_N \rangle$ is a single
many-electron configuration where the state on each site $i$ is $s_i$.
In the spin-$\frac{1}{2}$ model considered in Reference
\cite{Sandvik07a} $s_i=\pm 1$.  For the Hubbard model $s_i$
takes four possible values corresponding to either empty, spin-up,
spin-down, or doubly occupied states. $A_j$ are $D\times D$ matrices,
where $D$ is an adjustable parameter. As we study ground state
configurations containing spinon defect states that break
translational symmetry, we keep separate matrices $A_i(s_i)$ for each
lattice site. In the MPS-QMC method the matrix elements of the
matrices $A_i(s_i)$ are stochastically optimized to minimize the
expectation value of $H$.

The method may be summarized as follows: the matrix elements
$a^k_{ij}(s_k)$ are first initialized to hold random values. Matrices
are normalized so that their Frobenious norm is unity,
$\sqrt{\rm{Tr}(A A^T)}=1$. The expectation values of the energy and
derivatives of the energy with respect to $a^k_{ij}(s_k)$, $k=1\cdots
N$, are evaluated using MC sampling and a Metropolis acceptance
probability. Configurations are sampled according to the weight
$W^2(S)$ with
\begin{equation}
W(S)=\rm{Tr}[A_1(s_1) A_2(s_2) \cdots A_N(s_N)].
\label{weight}
\end{equation}
The MC updates used to sample the configurations $\{S\}$ interchange
electrons between neighboring sites, i.e.  $\{${\underbar{ }}
{\underbar{$\uparrow$}}$\}\rightleftarrows \{${\underbar{$\uparrow$}}
{\underbar{ }}$\}$. In such a process, two matrices in
Eq.~\ref{weight} are changed. The changed weight, $W(S^\prime)$, may
be efficiently generated by saving a series of ``right'' and ``left''
matrix products, and sequentially attempting updates\cite{Sandvik07a}
for sites $i=1\cdots N$.  The estimates of the derivatives are used to
optimize the initial random matrices using a stochastic optimization
scheme \cite{Sandvik07a}.  The calculation is divided into a series of
steps $k=0,1,2,\cdots$. At each step, the amplitude of the stochastic
noise is decreased, and the number of MC samples is increased to
provided increasingly accurate derivatives as the minimum energy is
approached.  The overall operation count of the MPS-QMC method scales
more favorably with the matrix size ($\propto D^3$) for periodic
systems than does traditional DMRG \cite{Sandvik07a}.
 \begin{table}
\small
\begin{center}
 \begin{tabular}{l|l|l|l|l}
 \hline
  $D$ & E & KE & PE &   $S_\sigma(\pi/2)$  \\
 \hline
  4 & -0.92942(1) &-1.46896(4) & 0.53954(4)  & 0.15115(3) \\
 8 & 0.940809(7) &-1.46732(4) & 0.52651(4) & 0.15280(3) \\
 16 & -0.942555(2) &-1.46824(4) &0.52568(4) & 0.15608(2) \\
 32 & -0.9426859(6) & -1.46818(3) & 0.52549(3) & 0.15647(4)  \\
 48 & -0.9427005(5)  & -1.46821(3) & 0.52551(3) & 0.15649(4) \\
 \hline
 ED\hspace*{-2mm} & -0.9427135 &-1.468223 & 0.525510 & 0.156520 \\
 \hline
 \end{tabular}
 \caption{Comparison of the ground state energy per site, kinetic
   energy per site, potential energy per site, and spin structure
   factor $S_\sigma(\pi/2)$ between MPS-QMC and exact diagonalization
   a 20 site ladder with periodic boundary conditions and $t_2/t_1=1$, $U=6$,
   and $V=1$. Statistical Monte Carlo sampling errors in the last
   digit are shown in parenthesis.}
 \label{mps-test2}
 \end{center}
\end{table}

Fig.~\ref{mps-test1} 
and Table \ref{mps-test2} compare the accuracy of MPS-QMC against
exact diagonalization for the energy, kinetic energy, potential energy,
and spin structure factor, defined as
\begin{equation}
S_\sigma(q)=\frac{1}{N}\sum_{jl}e^{iq(j-l)}\langle (n_{j\uparrow}-n_{j\downarrow})(n_{l\uparrow}-n_{l\downarrow})\rangle.
\end{equation}
Fig.~\ref{mps-test1} shows the relative error in the ground state
energy as a function of the number of sampling blocks for a 20-site
ladder with $t_1=t_2=1$, $U=6$ and $V=1$.  Each step $k$ of the
algorithm is further divided into a number $n$ of blocks consisting of
$m$ Monte Carlo updates each, with the matrices $A_i(S_i)$ updated
using the calculated derivatives after each block.  $n$ and $m$ are
increased as $n=k n_0$ and $m=k m_0$, with $n_0=25$ and $m_0=100$ in
Fig.~\ref{mps-test1}.  We find that larger matrix sizes are needed
here compared to those typically used for spin models; this is
expected because of the greater number of quantum states per site.
However, even with the much larger number of parameters (matrices are
also site-dependent in our calculations) the stochastic optimization
method performs well. Comparing with $N=20$ site exact results,
relative energy accuracy of order 10$^{-5}$ is achievable with
$D=48$. As shown in Table \ref{mps-test2}, local quantities such as
the kinetic energy (bond order) and potential energy converge quite
rapidly with $D$. As expected, long-range correlations require larger
$D$ for similar numerical accuracy. The 36-site calculations presented
in Section \ref{results} used $D=48$ matrices.

In order to treat the e-p terms in Eq.~\ref{ham}, we use the an
iterative self-consistent method\cite{Clay03a} to update the bond
distortions $\Delta_{i,j}$. The self-consistency equations for the
lattice are updated once every MC block following the calculation of
charge densities and bond orders. Self-consistent MPS-QMC results
were verified against exact self-consistent calculations.

\section{Finite-size scaling}
\label{appendixb}

For the results reported in Figs.~\ref{data} and
~\ref{vdata} we performed finite-size scaling from calculations
on 20, 28, 36,
and 60 site ladders.  The 20 site data was from exact lanczos
calculations and larger system data  from the MPS-QMC method
detailed in Appendix \ref{appendixa} using a matrix size of $D=48$.  
Here we give further details of our extrapolation procedure.

\begin{figure}
\centerline{\resizebox{3.2in}{!}{\includegraphics{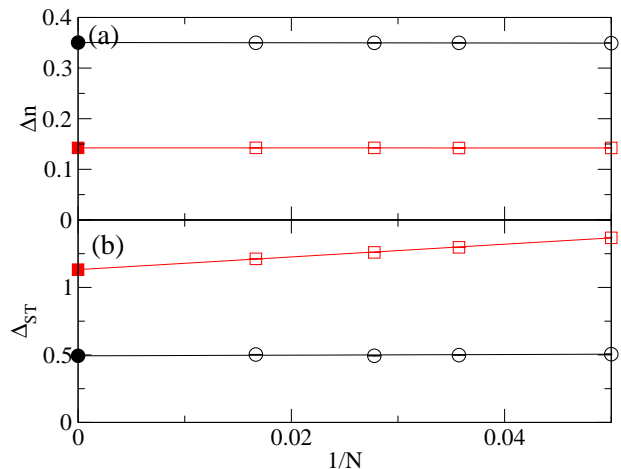}}}
\caption{(color online) Finite-size scaling of observables for $U=6$,
  $V=1$, and $\alpha=1.6$. Circles (squares) are for $t_2/t_1=0.2$
  ($t_2/t_1=1.6$). $N$ is the number of lattice sites. Filled symbols
show the extrapolated values of the observables. Panel (a) shows
the  charge disproportionation $\Delta n$ and panel (b) the singlet-triplet gap
  $\Delta_{\rm{ST}}$. Lines are least-squares fits to the points.}
\label{fss}
\end{figure}

Because the properties of quasi-1D systems display an alternation with
$N$ as the chain length increases, we have chosen only ladders with
$4n+2$ electrons.  These systems also have closed shells in the
noninteracting limit.  The PEC ground state further requires that $N$
be a multiple of four.  Fig.~\ref{fss}(a) shows the finite-size
extrapolation for the charge disproportionation $\Delta n$ for both
small and large $t_2/t_1$. $\Delta n$ as well as $r_4$ and $E_s$ (not
shown here) change only by a small amount from 20 to 60
sites. Fig.~\ref{fss}(b) shows the finite-size scaling of the
singlet-triplet gap $\Delta_{\rm{ST}}$. Errors in the extrapolation of
$\Delta_{\rm{ST}}$ are  larger than those in $\Delta n$, $r_4$,
and $E_s$, because the gap is calculated from the
energy difference of separate $S=0$ and $S=1$ calculations, each of
which have different lattice configurations that are optimized
self-consistently. This accounts for the larger error bars in 
Fig.~\ref{data}(d) compared to Fig.~\ref{data}(a)-(c).

\end{document}